\documentclass[aps,prd,preprintnumbers,superscriptaddress,nofootinbib]{revtex4-1}
\usepackage{amsmath}
\usepackage{amssymb}
\usepackage{amsthm}
\usepackage{todonotes}
\usepackage{mathtools}
\usepackage{mathrsfs}
\usepackage{bm}
\usepackage{slashed} 
\usepackage{graphicx}
\usepackage{multirow}
\usepackage{tikz}
\usepackage[caption=false]{subfig}
\usepackage{relsize}	
\usepackage{array}
\usepackage{float}
\usepackage{color}
\usepackage{xcolor}
\usepackage{soul}
\usepackage{verbatim} 

\allowdisplaybreaks

\newcommand{\beq}{\begin{equation}}
\newcommand{\eeq}{\end{equation}}
\newcommand{\be}{\begin{eqnarray}}
\newcommand{\ee}{\end{eqnarray}}

\newcommand{\Q}{\textsf{\scriptsize Q}}
\newcommand{\bsq}{{\boldsymbol q}^{\perp}}
\newcommand{\bsqq}{{\boldsymbol q}^{\perp 2}}

\newcommand{\bsk}{{\boldsymbol k}^{\perp}}
\newcommand{\bska}{{\boldsymbol\kappa}^\perp}
\newcommand{\bskapr}{{\boldsymbol\kappa}^{\prime \perp}}
\newcommand{\bskasq}{{\boldsymbol\kappa}^{\perp2}}

\newcommand{\bsb}{{\boldsymbol b}^{\perp}} 
\newcommand{\bsbb}{{\boldsymbol b}^{\perp2}} 
 
\newcommand{\bsP}{{\boldsymbol P}^{\perp}} 
 
\newcommand{\bsp}{{\boldsymbol p}^{\perp}} 
\newcommand{\bspp}{{\boldsymbol p}^{\perp 2}}

\newcommand{\es}{&=&}
\newcommand{\ps}{&+&}

\newcommand{\nn}{\nonumber}
\newcommand{\nnn}{\nonumber\\}

\usepackage{booktabs}
\AtBeginDocument{
\heavyrulewidth=.08em
\lightrulewidth=.05em
\cmidrulewidth=.03em
\belowrulesep=.65ex
\belowbottomsep=0pt
\aboverulesep=.4ex
\abovetopsep=0pt
\cmidrulesep=\doublerulesep
\cmidrulekern=.5em
\defaultaddspace=.5em
}

\begin{document}
\title{Gravitational form factors and mechanical properties of a quark at one loop in light-front Hamiltonian QCD }

\author{Jai More}
\email{jai.more@iitb.ac.in} \affiliation{ Department of Physics,
Indian Institute of Technology Bombay,Powai, Mumbai 400076,
India}

\author{Asmita Mukherjee}
\email{asmita@phy.iitb.ac.in} \affiliation{ Department of Physics,
Indian Institute of Technology Bombay, Powai, Mumbai 400076,
India}

\author{Sreeraj Nair}
\email{sreeraj@impcas.ac.cn} \affiliation{Institute of Modern Physics, Chinese Academy of Sciences, Lanzhou 730000, China}
\affiliation{School of Nuclear Science and Technology, University of Chinese Academy of Sciences, Beijing 100049, China}
\affiliation{CAS Key Laboratory of High Precision Nuclear Spectroscopy, Institute of Modern Physics, Chinese Academy of Sciences, Lanzhou 730000, China}

\author{Sudeep Saha}
\email{sudeepsaha@iitb.ac.in} \affiliation{ Department of Physics,
Indian Institute of Technology Bombay, Powai, Mumbai 400076,
India}

\date{\today}

\begin{abstract}
We calculate the gravitational form factors (GFFs) and pressure, shear and energy distributions for a quark state dressed with a gluon at one loop in QCD. We use the light-front Hamiltonian approach. In the light-front gauge, we use a two-component formalism to eliminate the constrained fields. The state may be thought of as a perturbative model for a relativistic spin $1/2$ composite system having a gluonic degree of freedom. We compare the results with model calculations for a nucleon.     

\end{abstract}

\maketitle

\section{Introduction}
 One of the major areas of interest in hadron physics and QCD in recent days is to understand the mechanical properties like mass, angular momentum and pressure distribution inside the nucleon in terms of quarks and gluons \cite{Polyakov:2018zvc,Lorce:2018egm,PhysRevD.99.094026}. The structure of the nucleon can be studied by bombarding it with a high energy probe like an electron. The mechanical properties of the nucleon are encoded in the gravitational form factors (GFFs), which are the form factors of the energy-momentum tensor.  The GFFs are functions of the square of the momentum transfer in the process ($Q^2$). They give information on how matter couples to gravity. These GFFs are related to the generalized parton distributions (GPDs) 
that can be accessed in exclusive electron-proton scattering process, for example deeply virtual Compton scattering (DVCS). The extensive experimental programs at HERA \cite{H1:1999pji,H1:2001nez,ZEUS:2003pwh}, HERMES \cite{HERMES:2001bob}, COMPASS \cite{ATLAS:2018zdn} and JLAB \cite{CLAS:2001wjj,CLAS:2020yqf} in the recent past have given valuable data as inputs for the extraction of the GPDs. The upcoming electron-ion collider (EIC) will also probe the GPDs as well as give inputs for  understanding of the mechanical properties of the nucleon.  There are  four GFFs of the proton, $A(Q^2), B(Q^2), C(Q^2)$ and $\overline{C}(Q^2)$. The GFFs $A(Q^2)$ and $B(Q^2)$ are related to the mass and angular momentum distributions of the proton. Conservation of the energy-momentum tensor constraints the GFFs  $A(Q^2), \  B(Q^2)$, and $\overline{C}(Q^2)$; however, $C(Q^2)$, also known as the D-term, is not related to any Poincare generator and is unconstrained by such conservation laws. This form factor, also known as the D-term, contributes to the DVCS process  when the skewness $\xi$ is non-zero, or when there is a longitudinal momentum transfer from the initial state proton to the final state proton. The pressure distribution inside the proton is related to the D-term. The pressure distribution is usually defined in the Breit frame, and it is subject to  relativistic corrections.  In Ref.\cite{Lorce:2018egm} the pressure distribution is defined in the infinite momentum frame, or equivalently in the light-front formalism \cite{Freese:2021czn}. As the transverse boosts in the light-front formalism are Galilean, one can describe the two-dimension pressure distribution of a relativistic system using overlaps of light-front wave functions (LFWFs) for the GFFs. The D-term is zero for a free fermion and $-1$ for a boson. For an interacting system, in order to ensure stability, the D-term has to be negative.  Recently the D-term has been extracted from the JLab data; the pressure distribution inside the nucleon obtained from fits of JLab data is found to be repulsive near the center and attractive towards the periphery of the nucleus \cite{Burkert:2018bqq}. At the core, the pressure is even higher than that inside a neutron star, which is the most dense object in the universe. The JLab result generated a lot of interest in this field and in the recent past, quite a lot of theoretical studies have been performed on the shear and pressure distributions in the nucleon in different models for example in the Bag model \cite{Neubelt:2019sou}, chiral quark soliton model \cite{Goeke:2007fp,Wakamatsu:2007uc,Schweitzer:2002nm,Goeke:2007fq,Wakamatsu:2006dy}, AdS/QCD motivated diquark model \cite{Chakrabarti:2015lba} as well as a simple multipole model \cite{Lorce:2018egm}. Lattice results are also available \cite{Hagler:2003jd,Gockeler:2003jfa,LHPC:2010jcs,LHPC:2007blg,QCDSF-UKQCD:2007gdl,Deka:2013zha}. 

Most of these models are phenomenological models for the proton and they do not incorporate any gluonic  degree of freedom. However, the D-term in particular, which is related to the pressure distribution inside the nucleon; involves the light-cone transverse  component of the energy-momentum tensor, that  depends on the quark-gluon interactions and hence the gluons are expected to play a major role here.  While  it is non-trivial to incorporate gluons in models for the nucleon,  which is a bound state, it is often interesting to replace the hadron state by  a simpler relativistic spin-$1/2$ state like a quark dressed with a gluon at one loop in QCD. This may be thought of as a perturbative model having a gluonic degree of freedom. The dressed quark state can be expanded in Fock space in terms of multi-parton LFWFs. These LFWFs can be calculated analytically using the light-front Hamiltonian. Upto one loop, this model incorporates the full quark-gluon interaction. 
The GPDs can be expressed in terms of overlaps of these LFWFs. Previously the GPDs as well as the DVCS amplitudes  have been calculated in a similar QED model \cite{PhysRevD.71.014038}. The  Wigner functions and spin distributions have been investigated in \cite{More:2017zqq,Mukherjee:2014nya}. In this work, we investigate the GFFs and the pressure distributions for a quark dressed with a gluon at one loop. The twist two GPDs as well as the GFFs $A(Q^2), \  B(Q^2)$ require a calculation of the `good' light-cone component of the energy-momentum tensor, whereas the remaining two GFFs, $C(Q^2)$ and $\overline{C} (Q^2)$ requires  a calculation of some of the `bad' components, which include the interaction terms. In order to calculate the matrix elements of these terms, we use the two-component formulation of light-front QCD developed in \cite{Zhang:1993dd} in light-cone gauge, $A^+=0$, where $A^\mu$ is the gluon field. The `bad' component of the fermion field in this case is constrained  and can be eliminated from the operator expressions, which would then be written only in terms of the dynamical quark and gluon fields.  Matrix elements of these interaction terms in the operator structures would involve particle number changing off-diagonal contributions, when expressed in terms of the multi-parton LFWFs. There would be diagonal contributions as well, that does not change the particle number.  In this work, we keep all terms upto $O(g^2)$, where $g$ denotes the quark-gluon coupling. A similar perturbative model was used in \cite{Metz:2021lqv} in QED, where the D-term  and the pressure and shear distribution were calculated for an electron dressed with a photon at one loop in QED using the Feynman diagram approach and in the Breit frame.  

The plan of the paper is as follows.  In section \ref{model} we introduce the gravitational form factors in a light-front dressed quark model, in section  \ref{Dterm} we discuss the $D$- term and the pressure distributions. Finally in section \ref{con} we present the summary and conclusion. Details of the calculation and useful formulas are given in the appendices.

\section{Gravitational form factors in a Light-front dressed quark model}
\label{model}
\subsection{Light-front wave function}
In this section, we outline the formalism used to calculate the gravitational form factors (GFFs) for a dressed quark state, which we call dressed quark model (DQM).  Instead of a proton state, we take  a quark dressed with a gluon.  This is a composite spin-$1/2$ state, which in the light-front (LF) formalism is treated fully relativistically. To list a few merits of this formalism : 
\begin{itemize}
    \item Due to the presence of gluon dressing, the model employs a  gluonic degree of freedom \cite{Harindranath:1996sj,Harindranath:1998pd}.
    \item The dressed quark state can be expanded in terms of light-front wavefunctions (LFWFs).  Although the LFWF of a bound state, like a proton, cannot be calculated analytically, the LFWF for a dressed quark can be calculated analytically in perturbation theory \cite{PhysRevD.63.045006}.
    \item  One can write the LFWFs in terms of relative momenta that are frame independent \cite{Brodsky:1997de}. Thus LFWFs are boost invariant.
    \end{itemize}
 
In LF Hamiltonian formalism, the dressed quark state can be expanded in Fock space, where the contributions come from a  single quark state, a quark and a gluon state, quark and two gluons state, and so on. However, we truncate the Fock space expansion upto the two-particle sector. Such truncation, in LF framework,  is boost invariant. The dressed quark state is written as   \cite{Harindranath:1998pd, Harindranath:1998fm}
\be \label{state}
|P,\lambda \rangle 
\es \psi_1 (P, \lambda) b^{\dagger}_{\lambda}(P)|0 \rangle + \sum_{\lambda_1, \lambda_2}\int[k_1][k_2]   \sqrt{2(2\pi)^3P^+} \delta^3 (P-k_1-k_2)\ \psi_2 (P,\lambda|k_1,\lambda_1;k_2,\lambda_2) b^{\dagger}_{\lambda_1}(k_1) a^{\dagger}_{\lambda_2}(k_2)|0\rangle,  \nnn 
\text{where}~~ [k]\es \frac{dk^+ d^2\bsk} {\sqrt{2(2\pi)^3{k^{+}}}}.
\ee
In Eq. (\ref{state}), $\psi_1(P, \lambda)$ in the first term, corresponds to a single particle state with momentum (helicity) $P$ ($\lambda$) and is  also responsible for the wavefunction normalization. The two-particle LFWF, $\psi_2(P,\lambda|k_1,\lambda_1;k_2,\lambda_2)$ is related to the probability amplitude of finding two particles namely a quark and a gluon with helicity $\lambda_1$ and $\lambda_2$ inside the dressed quark state. 
$b^\dagger$ and $a^\dagger$ correspond to the creation operator of quark and gluon respectively. 

In LF Hamiltonian framework \cite{Zhang:1993dd}, one uses the constraint equations in the light-cone gauge to eliminate the redundant degree of freedom and express the fields in terms of physically independent degrees of freedom. The quark fields are decomposed as \cite{Zhang:1993dd}
\beq
\psi=\psi_+\ +\ \psi_-,
\eeq
where $\psi_{\pm}=\Lambda_{\pm} \psi$ and $\Lambda_{\pm}$ are the projection operators. 

We use the  two-component framework developed in \cite{Zhang:1993dd}, where using a suitable  representation of the gamma matrices one can write:
\begin{align}
 	\psi_+= \begin{bmatrix}
 		\xi\\0
 	\end{bmatrix}, ~~~~\psi_-=\begin{bmatrix}
 	0\\ \eta
 \end{bmatrix}, 
 \end{align}
where, the two-component quark fields are given by 
\be
\xi(y) \es \sum_{\lambda}\chi_{\lambda}\int \frac{[k]}{\sqrt{2(2\pi)^3}}[b_{\lambda}(k)e^{-ik\cdot y}+d^{\dagger}_{-\lambda}(k)e^{ik\cdot y}],
\\
\eta(y) \es \left(\frac{1}{i\partial^+}\right)\left[\sigma^{\perp}\cdot\left(i\partial^{\perp}+g A^{\perp}(y)\right)+im\right]\xi(y),
\ee
and the dynamical components of the gluon field are given by 
\begin{align}
	A^{\perp}(y)= \sum_{\lambda} \int \frac{[k]}{\sqrt{2(2\pi)^3k^+}}[{\bf\epsilon}^{\perp}_{\lambda}a_{\lambda}(k)e^{-i k \cdot y}+ {\bf \epsilon}^{\perp*}_{\lambda}a^{\dagger}_{\lambda}(k)e^{i k \cdot y}].
\end{align}
Here we have suppressed the color indices. 

The LFWFs are written in terms of relative momenta \cite{Harindranath:1998pd, Harindranath:1998pc} and hence are independent of the momentum of the bound state. The Jacobi momenta $x_i$, $\bska_i$ are defined  such that they satisfy the relation $x_1+x_2=1$ and $\bska_1+\bska_2=0$.
\be
k_i^+=x_iP^+,~~~~ \bsk_i=\bska_i+x_i \bsP,
\ee 
where $x_i$ is the longitudinal momentum fraction for the quark or gluon , inside the two-particle LFWF.
The boost invariant two-particle LFWF can be written as,
\be\label{BILFWF}
\phi^{\lambda a}_{\lambda_1,\lambda_2}(x_i,\bska_i)\es \bigg[\frac{x(1-x)}{\bskasq+m^2(1-x)^2}\bigg]\frac{g}{\sqrt{2(2\pi)^3}}\frac{T^a}{\sqrt{1-x}} \chi_{\lambda_1}^{\dagger}\nnn
&\times&\bigg[\frac{-2(\bska\cdot \epsilon_{\lambda_2}^{\perp*})}{1-x}-\frac{1}{x}(\tilde{\sigma}^{\perp}\cdot\bska)(\tilde{\sigma}^{\perp}\cdot \epsilon_{\lambda_2}^{\perp*})+im(\tilde{\sigma}^{\perp}\cdot \epsilon_{\lambda_2}^{\perp*})\frac{1-x}{x}\bigg]\chi_{\lambda} \psi_1^{\lambda}
\ee
where, $\phi^{\lambda a}_{\lambda_1,\lambda_2}(x_i,\bska_i )= \sqrt{P^+} \psi_2 (P,\lambda|k_1,\lambda_1;k_2,\lambda_2)$,  $g$ is the quark-gluon  coupling. $T^a$ and ${\boldsymbol \epsilon}_{\lambda_2}^\perp$ are colour SU(3) matrices and polarization vector of gluon. The quark mass and the two-component spinor for the quark are denoted by $m$ and $\chi_\lambda$ respectively, $\lambda=1,2$ correspond to helicity up/down. We have used the notation  $\tilde{\sigma}_1=\sigma_2$ and $\tilde{\sigma}_2=-\sigma_1$ \cite{Harindranath:2001rc}. 
It is customary to define the four momenta in light-front as 
\be
P^\mu \es(P^+,\bsP,P^-). 
\ee
We choose Drell-Yan frame (DYF), so the longitudinal momentum transfer $q^+=0$.  Thus, the initial and the final state four momenta will be
\be\label{initialmom}
P^{\mu} \es\bigg(P^+, {\bf0}^{\perp}, \ \frac{M^2}{P^+}\bigg),\\
\label{finalmom}
P^{\prime\mu}\es\bigg(P^+,\ \bsq,\ \frac{\bsqq  + M^2}{P^+}\bigg), 
\ee
and the invariant momentum transfer
\be\label{momtranfer}
q^\mu\es(P^{\prime}-P)^\mu=\bigg(0, \ \bsq,   \frac{\bsqq}{P^+}\bigg).
\ee
In our frame,  $\bsP=0$ and $q^2=-\bsqq$.

\subsection{Energy-momentum tensor  and extraction of gravitational form factors}
We start by writing QCD Lagrangian
\be
\mathcal{L}_{QCD} \es  \frac{1}{2}\overline{\psi}\left(i \gamma_{\mu}D ^{\mu}-m\right)\psi-  \frac{1}{4} F^{\mu \nu}_a F^a_{\mu \nu},  \ee
where
the covariant derivative $
iD^{\mu} = i\overleftrightarrow{\partial}^\mu+gA^{\mu} $ and  
$\alpha (i\overleftrightarrow{\partial}^\mu)\beta
= \frac{i}{2}\alpha\left(\partial^{\mu}\beta\right) -\frac{i}{2}\left(\partial^{\mu}\alpha\right) \beta$. The field strength tensor for non-Abelian gauge theory is
\be 
F^{\mu \nu}_a = \partial^{\mu} A^{\nu}_a - \partial^{\nu} A^{\mu}_a +  g \ f^{abc} A^{\mu}_b A^\nu_c.
\ee
$\psi$ and $A^{\mu}$ is the fermion and boson field respectively.
The above Lagrangian can be used to obtain the gauge invariant symmetric  energy-momentum tensor \cite{Harindranath:1997kk}
\be
\theta^{\mu \nu} = \frac{1}{2}\overline{\psi}\ i\left[\gamma^{\mu}D^{\nu}+\gamma^{\nu}D^{\mu}\right]\psi - F^{\mu \lambda a}F_{\lambda a}^{\nu} + \frac{1}{4} g^{\mu \nu} \left( F_{\lambda \sigma a}\right)^2 - g^{\mu \nu} \overline{\psi} \left(
i\gamma^{\lambda}D_{\lambda} -m
\right)
\psi. \label{emtqcd}
\ee
The last term in Eq. \ref{emtqcd} goes to zero when one writes Lagrangian equation of motion of fermions. We are interested only in the fermionic part of the energy-momentum tensor for this work. So, Eq. \ref{emtqcd} reduces to
\be
\theta^{\mu \nu}_\Q = \frac{1}{2}\overline{\psi}
i\left[\gamma^{\mu}D^{\nu}+\gamma^{\nu}D^{\mu}\right]
\psi \label{emtqcdforquark}
\ee
The matrix element of totally symmetric energy-momentum tensor (EMT) encodes the information of GFFs. In order to calculate the matrix elements, we use the dressed quark state defined in Eq. \ref{state} and extract the GFFs from the quark part of the symmetric energy-momentum tensor $\theta^{\mu \nu}_Q$.

For a spin half system, the standard parameterization used to obtain the four GFFs from the symmetric energy-momentum tensor up to $O(q^2)$ is \cite{Ji:2012vj,Harindranath:2013goa}
\be
	\langle P^{\prime},S^{\prime}|	\theta^{\mu\nu}_i(0)|P,S \rangle \es\overline{U}(P^{\prime},S^{\prime})\bigg[-B_i(q^2)\frac{\overline{P}^{\mu}\ \overline{P}^{ \nu}}{M}+\left(A_i(q^2)+B_i(q^2)\right)\frac{1}{2}(\gamma^{\mu}\overline{P}^{\nu}+\gamma^{\nu}\overline{P}^{\mu})\nnn
	\ps C_i(q^2)\frac{q^{\mu}q^{\nu}-q^2g^{\mu\nu}}{M}+\overline{C}_i(q^2)M\ g^{\mu\nu}\bigg]U(P,S), 
	\label{e36}
\label{FF}
\ee
where $\overline{P}^{\mu}=\frac{1}{2}(P^{\prime}+P)^{\mu}$ is the average nucleon four momentum. $\overline{U}(P',S')$, $U(P,S)$  are the Dirac spinors and $M$ is the mass of the target state, $i \equiv(Q,G)$. $A_i$, $B_i$, $C_i$ and $\overline{C}_i$ are the quark or gluon form factors. In the Drell-Yan frame, as discussed above, there is no momentum transfer in the longitudinal direction and from Eq. (\ref{FF}), one can easily extract the form factors $A_\Q(q^2)$ and $B_\Q(q^2)$ from the diagonal component of the energy-momentum tensor $\theta^{++}_{\Q}$. $A_i(q^2) \ (B_i(q^2))$ is obtained from the sum of the helicity conserving (flip) states. The outline of the calculation is briefly described below. 
The matrix element that one needs to calculate can be written in the compact form as
\be \label{matrixelement}
\mathcal{M}^{\mu \nu }_{SS'} = \frac{1}{2}\left[\langle P'  ,S'|	\theta^{\mu \nu }_\Q(0)|P,S \rangle \right]
\ee
where the Lorentz indices $(\mu, \nu)\ \equiv \{+,-,1,2\}$, $(S, S') \equiv \{ \uparrow,\downarrow \}$  is the helicity of the initial and final state. $\uparrow(\downarrow)$ positive (negative) spin projection along $z-$ axis. Using Eq. \ref{matrixelement} we obtain
\be\label{rhsA}
 \mathcal{M}^{++}_{\uparrow \uparrow} + \mathcal{M}^{++}_{\downarrow \downarrow} \es 2\  (P^+)^2A_\Q(q^2), \\
 \label{rhsB}
\mathcal{M}^{++}_{\uparrow \downarrow} + \mathcal{M}^{++}_{\downarrow \uparrow} \es \frac{ i q^{(2)}}{M} \ (P^+)^2 B_\Q(q^2) .
\ee
The most apt way to extract GFFs $C_\Q(q^2)$ and $\overline{C}_\Q(q^2)$ is using transverse components $(1, 2)$ of the energy-momentum tensor
\be\label{rhsC}
\mathcal{M}^{11}_{\uparrow \downarrow} + \mathcal{M}^{11}_{\downarrow \uparrow} - \mathcal{M}^{22}_{\uparrow \downarrow} - \mathcal{M}^{22}_{\downarrow \uparrow} \es i\bigg[\frac{B_\Q(q^2)}{4M}- \frac{C_Q(q^2)}{M}\bigg]\left((q^{(1)})^2 \ q^{(2)}-(q^{(2)})^3\right), \\
\label{rhsCbar}
\mathcal{M}^{11}_{\uparrow \downarrow} + \mathcal{M}^{11}_{\downarrow \uparrow} + \mathcal{M}^{22}_{\uparrow \downarrow} + \mathcal{M}^{22}_{\downarrow \uparrow} \es
i\bigg[B_\Q(q^2)\frac{q^2}{4M}-C_\Q(q^2)\frac{3q^2}{M}+\overline{C}_\Q(q^2) 2M\bigg]q^{(2)}.
\ee
Here $q^{(i)}$ are the components of $\bf{q}^\perp$.  By knowing the form of $B_\Q(q^2)$ from Eq. \ref{rhsB} and from Eq. \ref{rhsC} we obtain the analytical expression for $C_\Q(q^2)$. We can now easily calculate the form of $\overline{C}_\Q(q^2)$. 
The final expression for the four independent GFFs are as follows:
\be\label{gffAexp}
A_\Q(q^2)\es 1+ \frac{g^2\ C_F}{2 \pi ^2}\left[\frac{11}{10} -\frac{4}{5}\left(1+\frac{2 m^2}{q^2}\right) \frac{f_2}{f_1}-\frac{1}{3} \log \left(\frac{\Lambda ^2}{m^2}\right)\right]\\
B_\Q(q^2)\es \frac{g^2 C_F}{12\pi^2}\ \frac{m^2}{q^2}\ \frac{f_2}{f_1},\label{gffBexp}\\
D_\Q(q^2)\es  \frac{5 g^2 C_F}{6\pi^2}\  \frac{m^2}{q^2} \ \bigg(1-f_1f_2\bigg)= 4 \ C_\Q(q^2),\label{gffDexp}\\
\overline{C}_\Q(q^2)\es \frac{g^2 C_F}{72\pi^2} \ \left(29-30 \ f_1 \ f_2+3~\mathrm{log}\bigg(\frac{\Lambda^2}{m^2}\bigg)\right),
\label{gffCbarexp}\ee
where
\be
f_1 :\es\frac{1}{2}\sqrt{1+\frac{4 m^2}{q^2}}, \\ f_2:\es\log\left(1+\frac{q^2\left(1+2f_1\right)}{2 m^2}\right)  
\ee
$C_F$ is the colour factor and $\Lambda$ is the ultra-violet cut-off.
The expression for all the form factors in terms of the overlap of LFWFs are given in Appendix \ref{appA}. 
The necessary forms of the integrals required for momentum integration in the transverse plane are also shown in Appendix  ~\ref{appB}. Then it is straightforward to obtain the compact form of the GFFs using Eq. (\ref{BILFWF}) and integrating over the Jacobi momenta $\bska$ and $x$. 
 The transverse momentum integrals of the GFFs $B_\Q(q^2)$ and $C_\Q(q^2)$ are convergent and hence are finite and so is the GFF $D_\Q(q^2)$ as it is four times $C_\Q(q^2)$. However, while calculating the transverse momentum integrals of the GFFs $A_\Q(q^2)$ and $\overline{C}_\Q(q^2)$ we encounter ultra-violet (UV) divergences which we regulate using the UV cut-off $\Lambda$. This is introduced as the upper limit of the transverse momentum integration. The longitudinal momentum integral involved in the calculation of the GFF $A_\Q(q^2)$ receives a contribution from the wave function normalization at $x=1$ \cite{Harindranath:1998pd, Harindranath:1998pc}, which contributes in the single particle sector upto the order we are interested in.  In this work, we included the contribution from the two-particle LFWF only, and we have included a cut-off at $x=1$ in the longitudinal momentum integration when calculating $A_\Q(q^2)$.

\subsection{Numerical Analysis: Gravitational Form Factors}

In this section, we show the numerical results for the four GFFs viz $A(q^2)$, $ B(q^2)$, $D (q^2)$ and $\overline{C}(q^2)$  shown in Eqs.~\ref{gffAexp} -\ref{gffCbarexp}. 
We now study the analytical results by fixing the parameters for all the GFFs:
the quark mass $m = 0.3 ~\mathrm{GeV}$, the coupling constant $g=1$, the color factor $C_F = 1$ and the UV cut-off $\Lambda = 10^3~\mathrm{GeV}$. We choose the cut-off such that $\Lambda>>M$, where $M$ is the mass of the dressed quark system. In this work, we take $M=m$.
\begin{figure}[h]
	\begin{minipage}{0.45\linewidth}
		\includegraphics[scale=0.3]{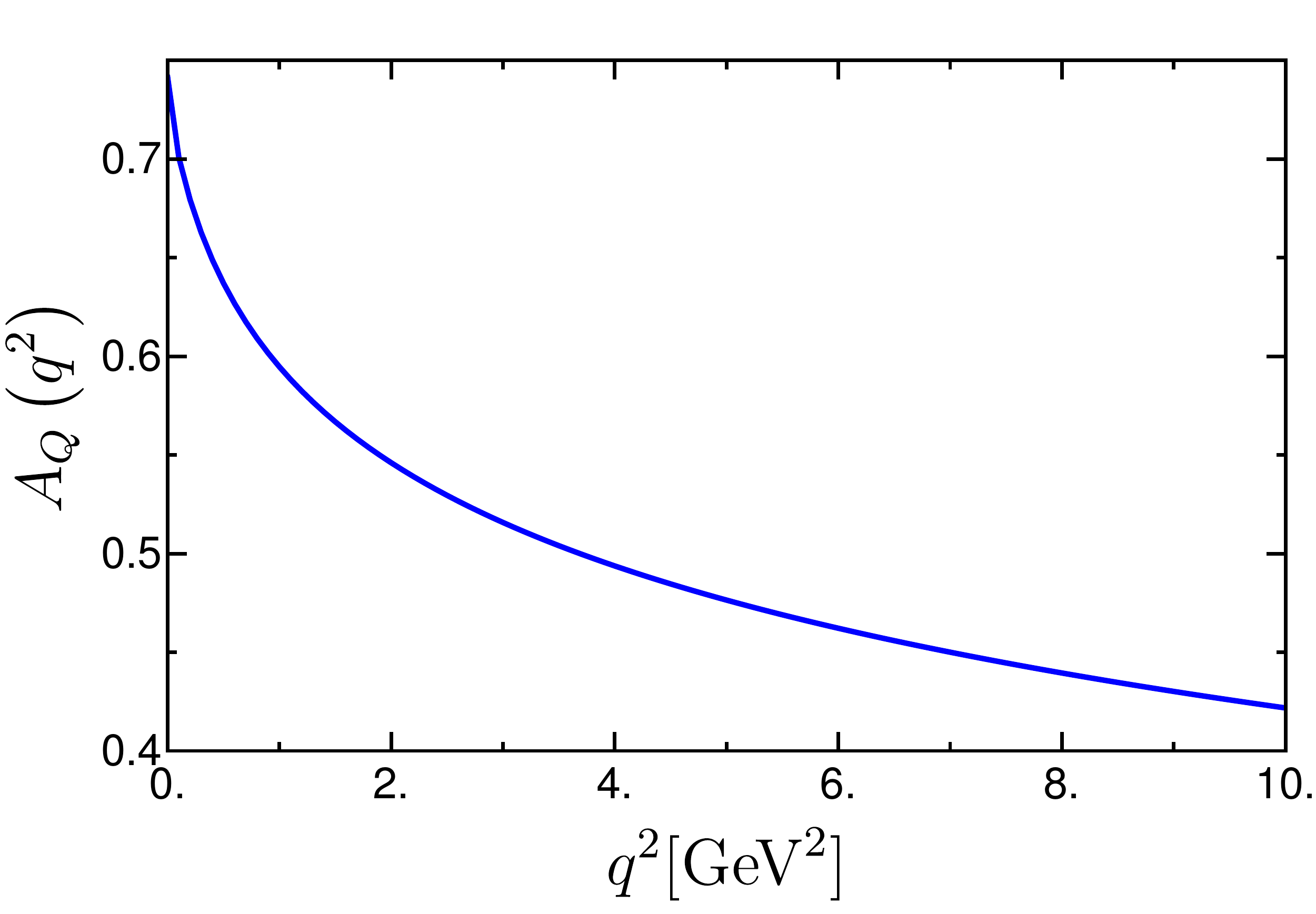}
	\end{minipage}
	\begin{minipage}{0.45\linewidth}
		\includegraphics[scale=0.3]{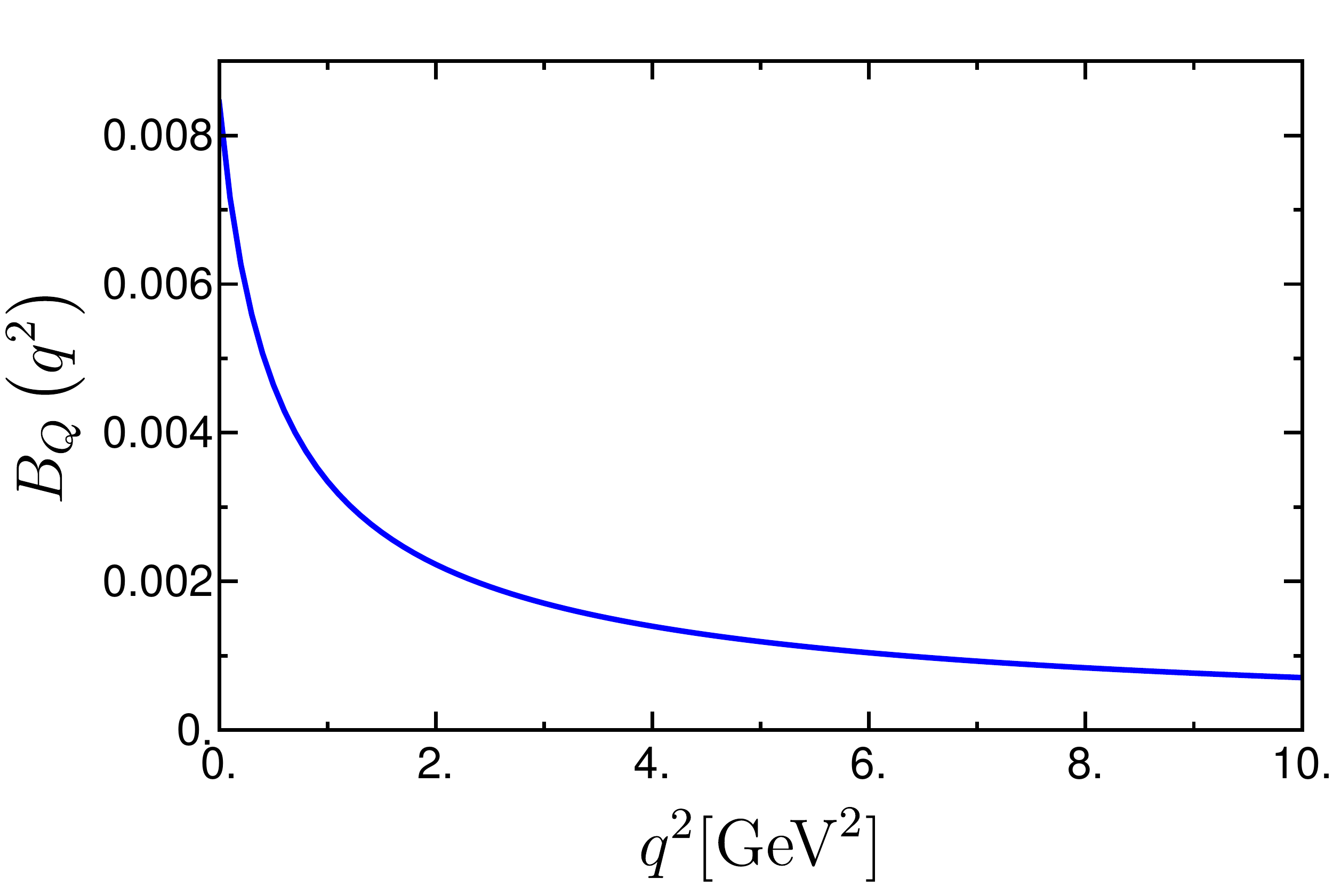}
	\end{minipage}  
	\caption{Plot of the GFFs $A_\Q(q^2)$ and $B_\Q(q^2)$ as function of $q^2$, here $m=0.3$ GeV and $g=1$.}
	\label{figgffAnB}
\end{figure}

In Fig.~\ref{figgffAnB}, we plot the GFFs $A_Q(q^2)$ and $B_Q(q^2)$ as functions of squared momentum transfer $q^2 $. The behavior of the form factors as $q^2 \rightarrow 0$  has received substantial interest in the literature~\cite{Lowdon_2017,
Hagler:2003jd,Dorati:2007bk,Lorce:2018egm,Kim:2012ts,Jung:2014jja}. The light-cone momenta of the parton is related to $A_i(0)$ and the total angular momenta of the parton is  $J_i(0) = \frac{1}{2}\left( A_i(0) + B_i(0)\right)$. 
The conservation of momentum and total angular momentum demands that $A(0) + B(0) =1$ and $J(0) = \frac{1}{2}\left( A(0) + B(0)\right)={1\over 2}$, summed over all quarks and gluons \cite{Teryaev:111207,Brodsky:2000ii,Teryaev:1999su}. 
 We observe in Fig.~\ref{figgffAnB} that both $A_Q(q^2)$ and $B_Q(q^2)$ are positive over the chosen 
$q^2$ range having their maxima at $q^2 = 0$. We see that $A_\Q(0) = 0.7412$ for $\Lambda = 10^3~\mathrm{GeV}$. 
The GFF $B_\Q(0) = 0.0084$ which can be interpreted as the value of the anomalous gravitomagnetic moment of the quark in the DQM. One can expect that the contribution coming from the gluon anomalous gravitomagnetic moment would be negative such the total $B(0)$ vanishes \cite{Brodsky:2000ii}. The calculation of the gluon part of the GFFs and verifying the sum rules is part of a future publication.  
$B_\Q(q^2)$ is related to the second moment of the helicity flip generalized parton distribution (GPD) $E(x,q^2)$ by the spin sum rule such that $B_i(q^2) = \int dx~ x E_i(x,q^2)$ ~\cite{Ji_1997}. In the case of QED the first moment of the GPD $E_e (x,0)$ corresponds to the Schwinger value for the anomalous magnetic moment $\int dx E_e (x,0) = F_2(0) = \frac{\alpha}{2\pi}$ \cite{PhysRevD.71.014038}. Our approach leads $B_e(0) = \frac{\alpha}{3\pi}$ for the case of electron, as expected  where $\alpha$ is the fine structure constant.
  \begin{figure}[htp!]
	\begin{minipage}{0.45\linewidth}
		\includegraphics[scale=0.3]{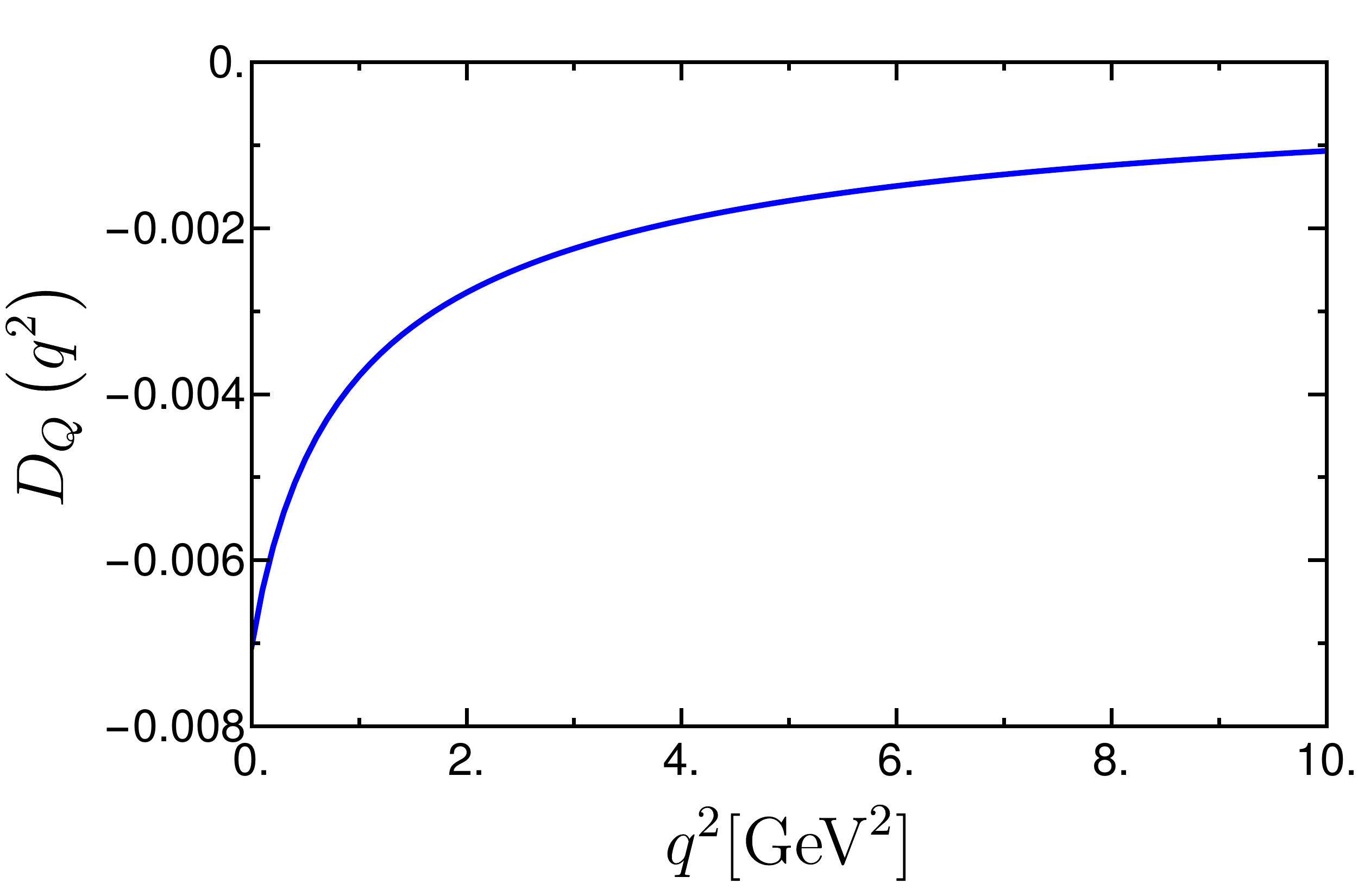}
	\end{minipage}
	\begin{minipage}{0.45\textwidth}
		\includegraphics[scale=0.3]{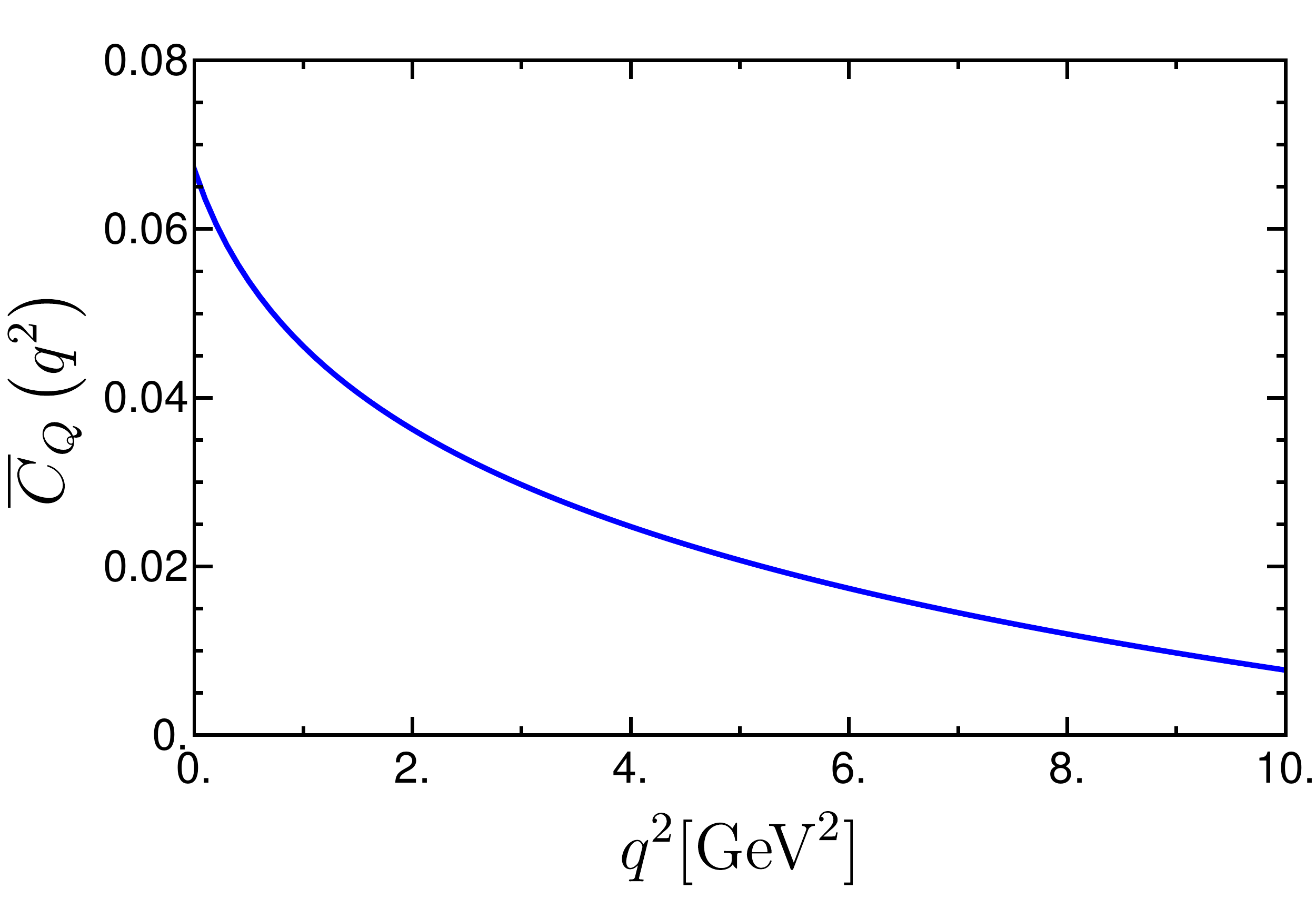}
		\label{gffDDQM}
	\end{minipage}  
	\caption{Plot of the GFFs $D(q^2)$ and $\overline{C}_\Q(q^2)$ as function of $q^2$, here $m=0.3$ GeV and $g=1$.}
		\label{figgffDnCbar}
\end{figure}

In Fig.~\ref{figgffDnCbar} shows our result for the GFFs $D_Q(q^2)$ and $\overline{C}_\Q(q^2)$ as function of $q^2$. We observe that $D_\Q(q^2)$ is negative which means that the DQM behaves like a bound system for the chosen set of parameters. The value of $D_\Q(0) = -\frac{5 g^2 \ C_F}{72\pi^2}$ which for our model parameter has numerical value of $D_\Q(0) = -0.007$. The GFF $\overline{C}_\Q(q^2)$ is found to be positive for the chosen range of $q^2$ and the value of $\overline{C}_\Q(0) = 0.067$. 
The sum of quark and gluon contribution for $\overline{C}(0) $ is constrained to be zero. Since we are getting a positive contribution from the quark,
 so we expect the gluon contribution to be negative in accordance with the constraint. The calculation of the gluon part is part of a future publication.  The sign of both $A_\Q(q^2)$ and $\overline{C}_\Q(q^2)$ is $\Lambda$ dependent.\\

\section{D-term and the Pressure Distribution}
\label{Dterm} 
The fundamental information like mass and spin of a particle is encrypted in the EMT which couples gravity to matter \cite{Pagels:1966zza,Polyakov:2002yz,Polyakov:2018zvc,Polyakov:2018exb,belitsky2005unraveling,diehl2003generalized}. The EMT, also offer internal mechanical properties via  \emph{D}-term which is called as the last unknown global property \cite{Polyakov:2002yz,Polyakov:2018zvc,Polyakov:2018exb}. As mentioned in the introduction, the \emph{D}-term is not related to any Poincare generators and is unconstrained.  The recent experimental extraction of the quark \emph{D}-term  resulted in the first experimental estimate of the pressure distribution inside the proton \cite{Burkert:2021ith}.

  \begin{figure}[htp!]
	\begin{minipage}{0.45\linewidth}
		\includegraphics[scale=0.35]{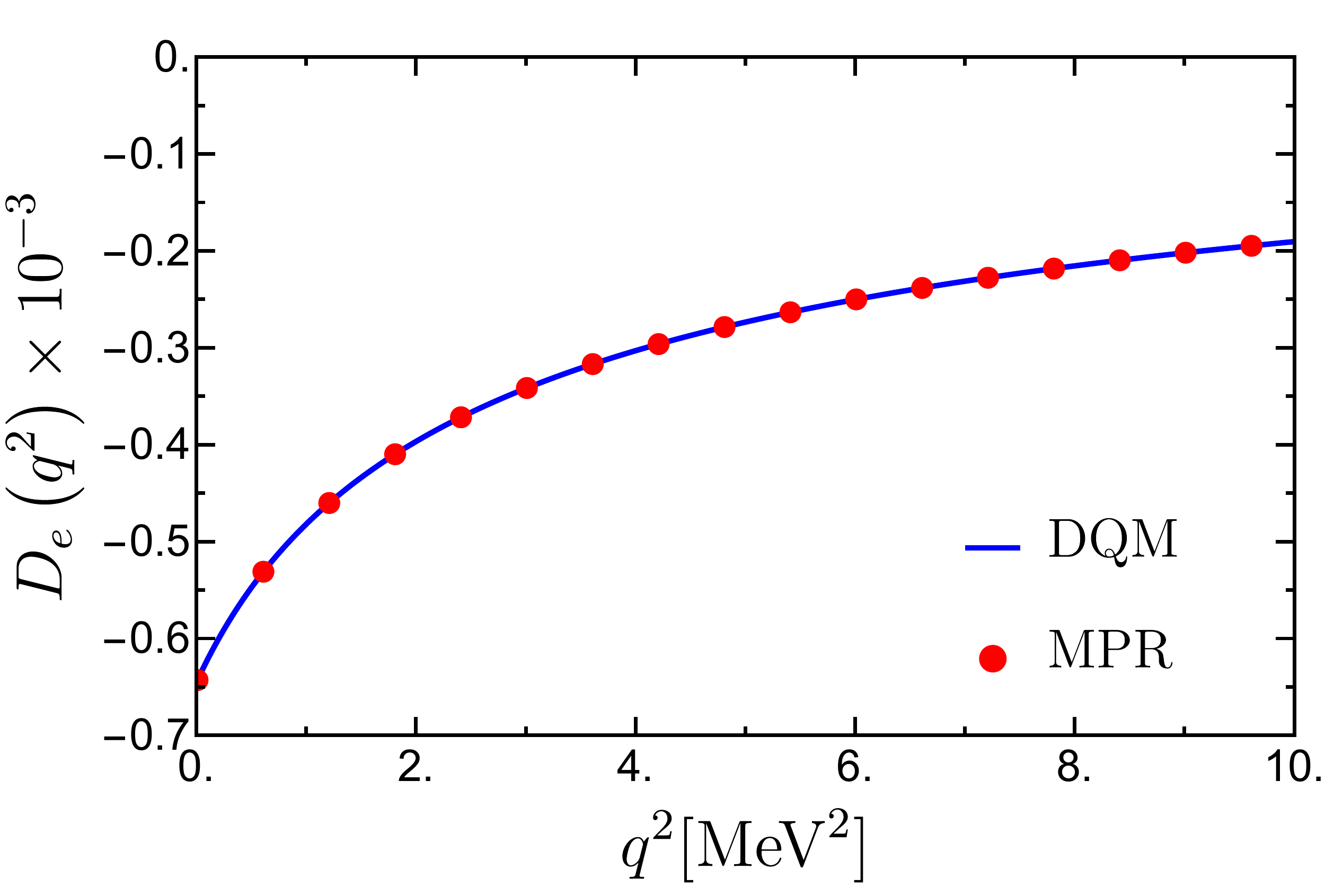}
	\end{minipage}
	\caption{Plot of electron GFF $D_e(q^2)$ as function of $q^2$, here we set $m=0.511$ MeV, $\alpha=\frac{1}{137}$. }
		\label{Dcompare}
\end{figure}
The electron gravitational form factor $D_e(q^2)$ was studied in \cite{Metz:2021lqv} using Feynman diagram approach. It was found in \cite{Metz:2021lqv} that the total D-term for a dressed electron is divergent at $q^2=0$, the divergent comes from the photon part of the QED energy-momentum tensor. A non-zero photon mass acts as a regulator for this divergence. In order to compare our result for $D_e(q^2)$ with Metz {\it et al.} (MPR) we use eq. [\ref{gffDexp}] and set the model parameters to QED domain such that, $m=0.511$ MeV and $g^2=4\pi \alpha$, where $\alpha$ is the fine structure constant. We observe that our result obtained in LFWF approach is in excellent agreement with MPR as seen in the plot in Fig. \ref{Dcompare}. 

\subsection{Pressure and shear forces for a quark dressed with a gluon}

We obtain the $D$-term from the transverse components of EMT that in turn is related to the pressure and shear distributions  \cite{Polyakov:2002yz,Polyakov:2018zvc,Polyakov:2018exb}.
These quantities can be well studied in the impact parameter space  by performing a Fourier transformation.  
We work in light-front framework, where the light-front time is $x^+$. At  $x^+=0$, the longitudinal LF coordinate $x^-$ can be integrated out from the EMT and hence we are left with two transverse coordinates \cite{Lorce:2018egm}. As the transverse boosts in the light-front framework are Galilean, one can define the pressure and shear distributions in 2D, which are free from relativistic corrections. The light-front definitions of 2D pressure and shear distributions are related to the 3D definitions in the Breit frame by Abel transformation \cite{Panteleeva:2021iip}. In the Drell-Yan frame, a two-dimensional Fourier transform of the $D$-term from the momentum space $\bsq$ to impact parameter space $\bsb$ \cite{Burkardt:2000za} gives the mechanical properties like the shear and pressure distributions in the transverse impact parameter space .
The expressions for pressure and shear distributions in two-dimensions \cite{Freese:2021czn} are 

\be
\label{Prefun}
p(\bsb)\es\frac{1}{2M \bsb}\frac{d}{d\bsb} \left[\bsb\ \frac{d}{d\bsb} D_\Q(\bsb)\right]-M \ \overline{C}_\Q(\bsb),\\
s(\bsb)\es-{\bsb\over M} \frac{d}{d\bsb}\left[ \frac{1}{\bsb} \frac{d}{d\bsb} D_\Q(\bsb)\right],
\ee

where
\be
\label{fgff}
F(\bsb)
\es  \frac{ 1}{(2\pi)^2}~\int d^2 \bsq \ e^{-i\bsq \bsb} \mathcal{F}(q^2) \nn \\
&=&\frac{1}{2\pi}\int_0^{\infty} d  \bsqq J_0\left( \bsq \bsb\right)\mathcal{F}(q^2),
\ee
where $\mathcal{F}=(A, B, C, \overline{C})$. $J_0$ is the Bessel function of the zeroth order and $\bsb$ represents the impact parameter. $M$ is the mass of the dressed quark state.

In order to study the spatial distribution,  we take  wave packet states  in position space centered at origin instead of plane waves. The most prevalent forms are {\it Gaussian wave packets} \cite{Diehl:2002he, Chakrabarti:2005zm}.  The dressed quark state which is confined in transverse momentum space and has definite longitudinal momentum can be written as  
\be
\frac{1}{16\pi^3}\int \frac{d^2 \bsp dp^+}{p^+}\phi\left( p\right) \mid p^+,\bsp,\lambda \rangle 
\ee
with $\phi(p)=p^+\ \delta(p^+-p_0^+)\ \phi\left(\bsp\right)$.
We choose a Gaussian shape for $\phi\left( \bsp\right)$ in transverse momentum :
\be\label{gaussian}
\phi\left(\bsp\right)
= e^{-\frac{\bspp}{2\Delta^2}}
\ee 
where $\Delta$ is the width of Gaussian. Since we choose to work in LF, the longitudinal and transverse momentum integration can be done independently.

\subsection{Numerical analysis: pressure and shear forces}

\begin{figure}[h]
	\begin{minipage}{0.45\linewidth}
		\includegraphics[scale=0.3]{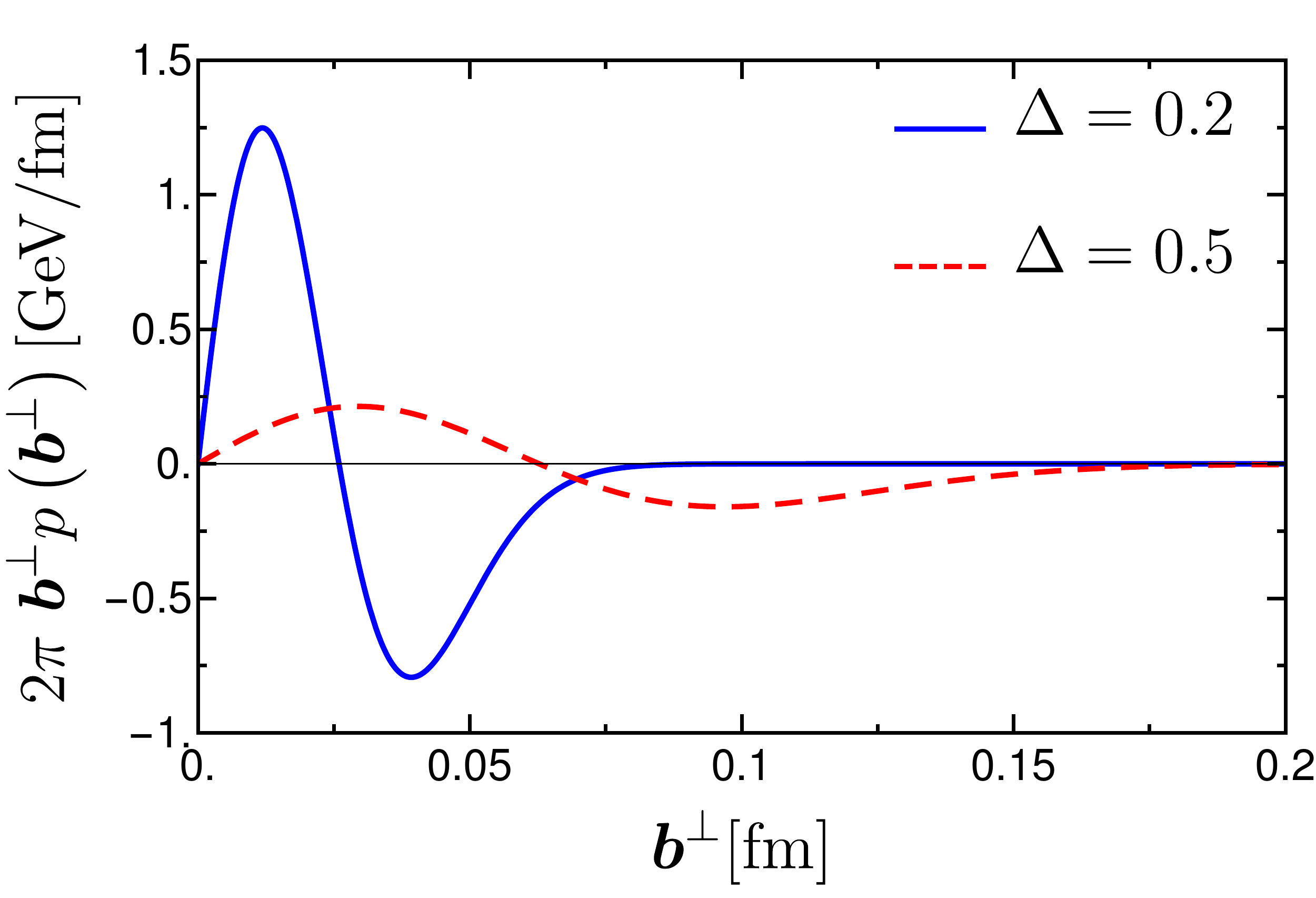}
	\end{minipage}
	\begin{minipage}{0.45\textwidth}
		\includegraphics[scale=0.3]{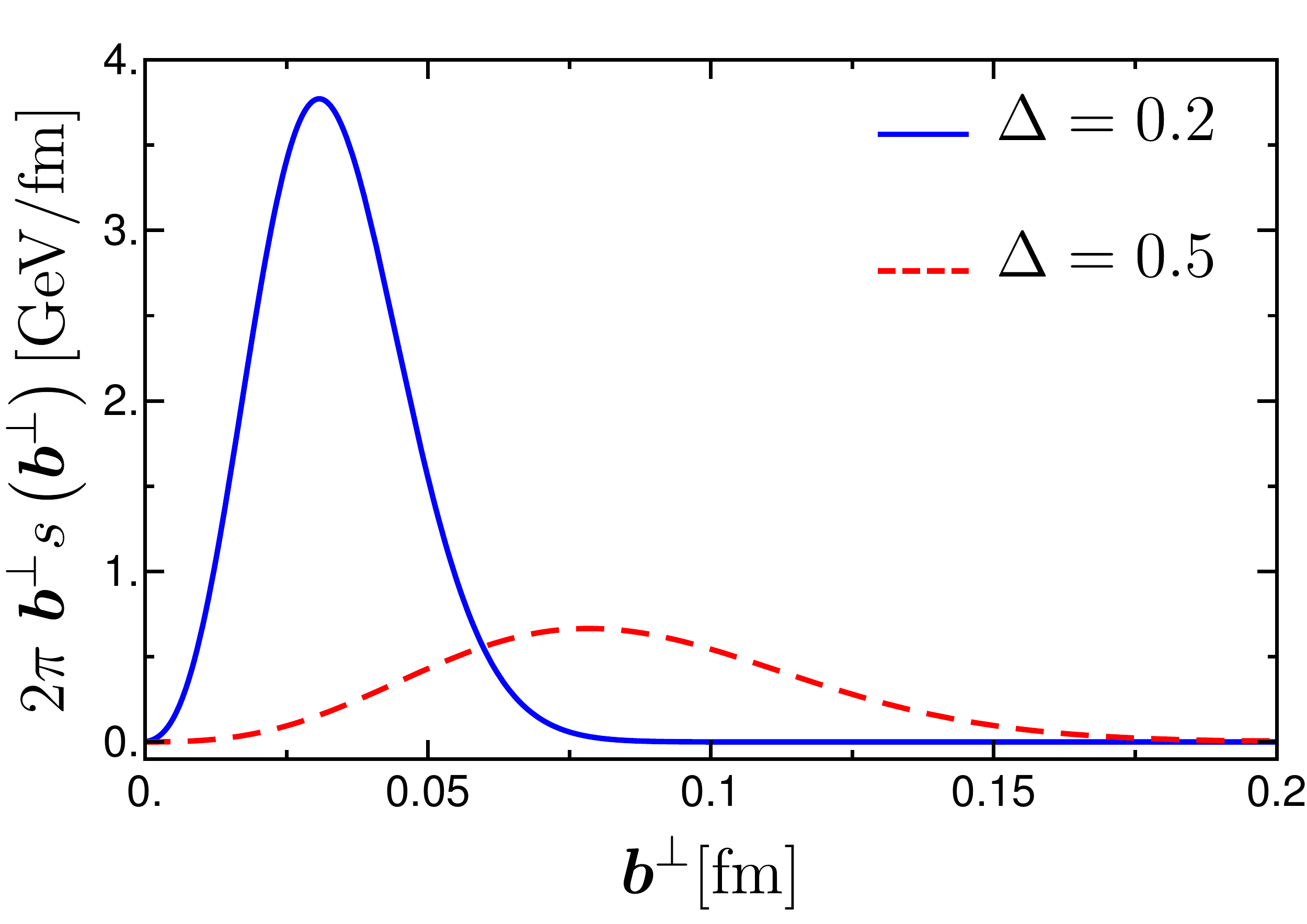}
					\end{minipage}  
	\caption{Plots of (left panel) the pressure distribution $2\pi \bsb p(\bsb)$ and (right panel) the shear force distribution $2\pi \bsb s(\bsb)$ as a function of $\bsb$.}
	\label{figPresuure}
\end{figure}

In Fig.~\ref{figPresuure} we have shown the plot for the pressure $2\pi \bsb \ p(\bsb)$ and shear $2\pi \bsb \ s(\bsb)$ distributions in the impact parameter space. We have chosen wave packets over the plane waves as it not only yields the Fourier transformed pressure in the impact 
parameter space \cite{Diehl:2002he} but also gives smooth plots for the distribution. The spread of the distribution depends on the width of the Gaussian $\Delta$. So, we study the dependence of $\Delta$ in Fig.~\ref{figPresuure} by plotting the distributions for two different values of $\Delta$.

 In our model, pressure distribution is in accordance 
with the von Laue condition \cite{Lane,Freese:2021czn} which is a stability condition such that \be\int_0^\infty d^2 \bsb ~p(\bsb) = 0.
\ee
The stability condition requires the presence of at-least one node in the pressure distribution which is seen in Fig.~\ref{figPresuure} (left panel). 
The pressure profile shows a central positive core which becomes negative towards the tail region. The repulsive central pressure in contrast with the confining pressure in the outer region maintains the stability of the system \cite{Burkert:2018bqq}. We observe that the shear distribution is 
non-negative in the b-space. This behavior of the shear force distributions is also observed in stable hydrostatic systems \cite{Polyakov:2018zvc}.
We also observe that as the width of the Gaussian wave packet, $\Delta$ increases the peak value of the distribution decreases and shifts away from the center in the impact parameter space. 

\begin{figure}[htp!]
	\begin{minipage}{0.45\linewidth}
		\includegraphics[scale=0.3]{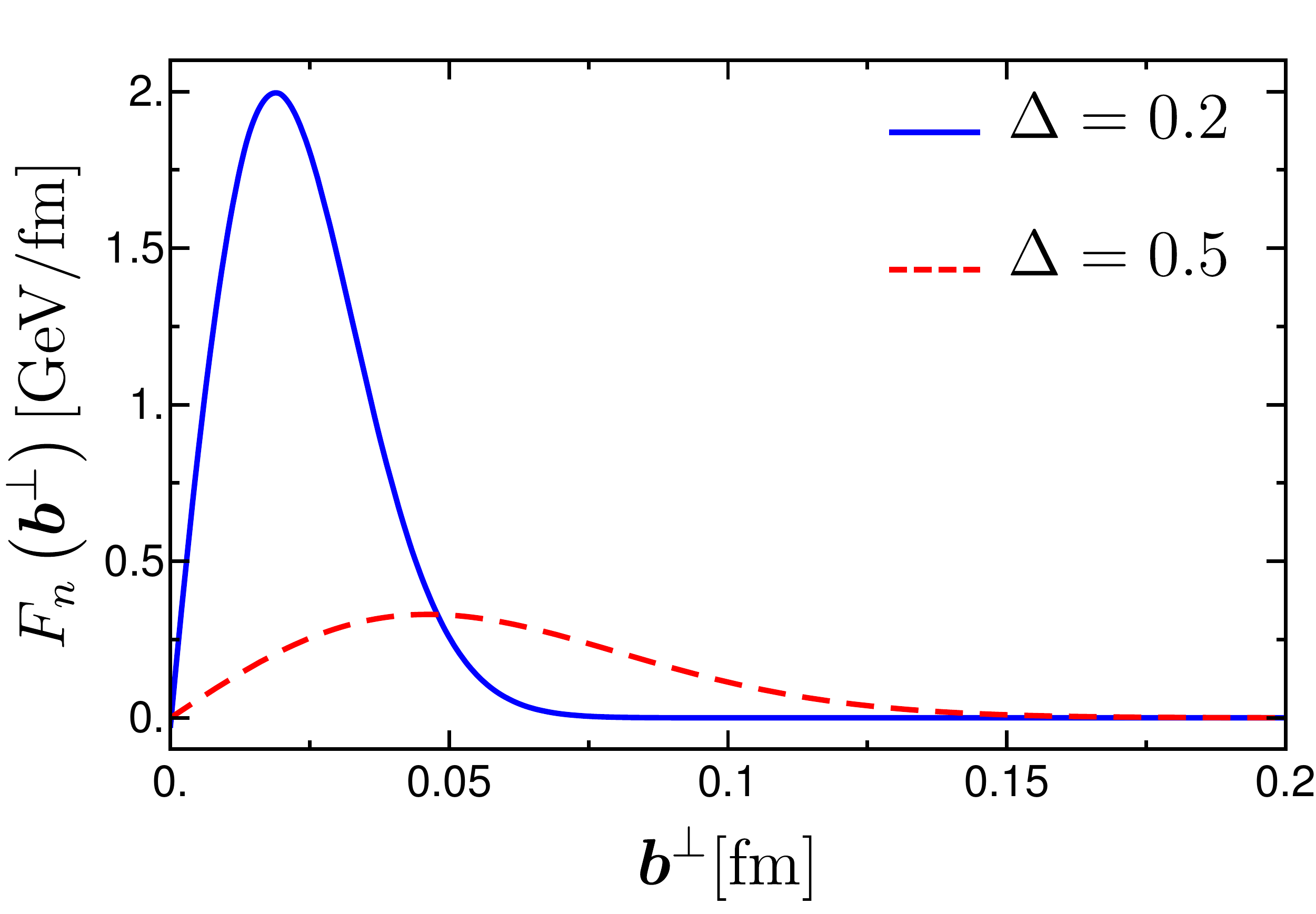}
	\end{minipage}
	\begin{minipage}{0.45\textwidth}
		\includegraphics[scale=0.3]{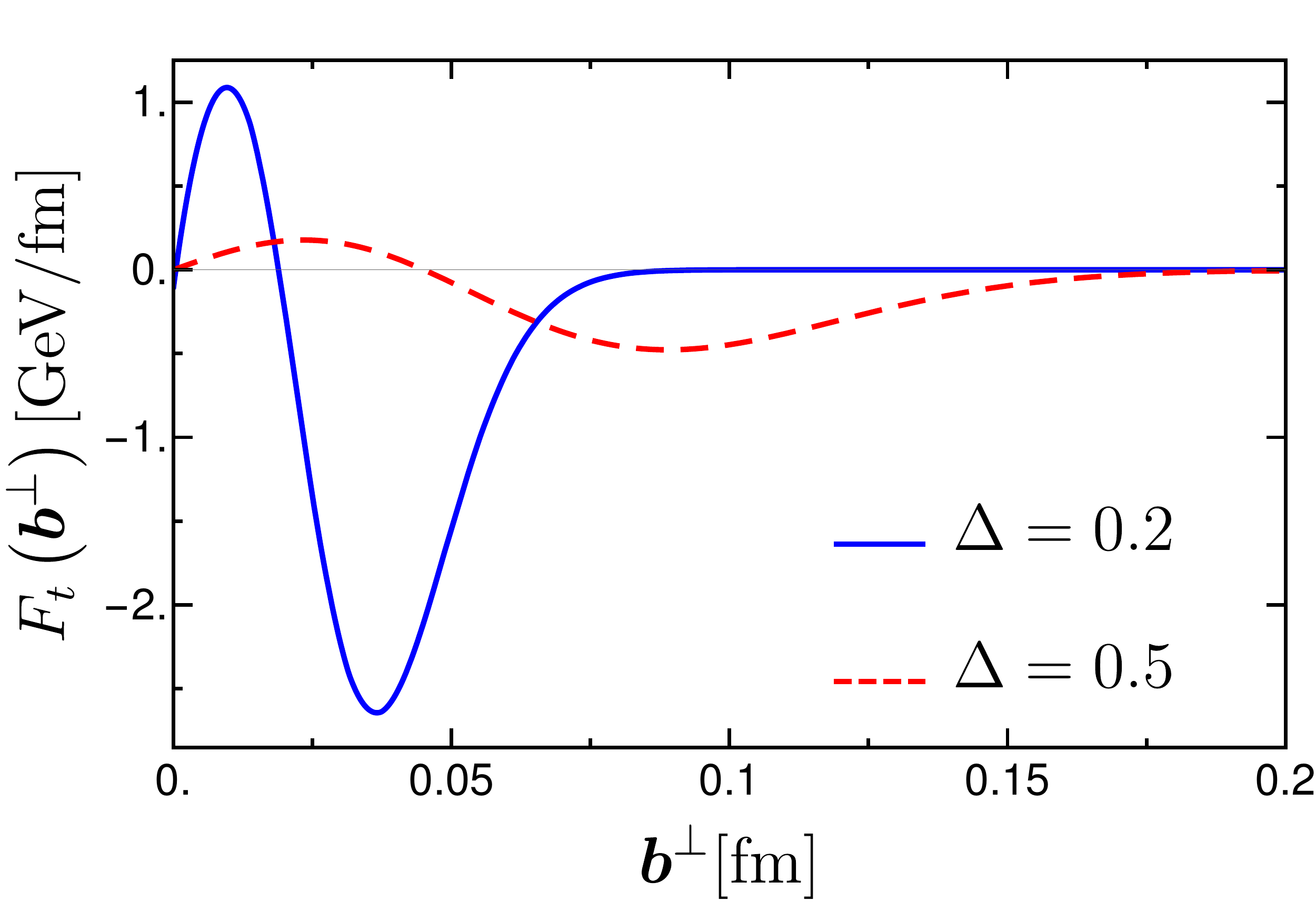}
		
	\end{minipage}  
	\caption{Plots of (left panel) the normal forces  $F_n(\bsb)$, and (right panel) the tangential forces $F_t(\bsb)$ as a function of $\bsb$.}
	\label{figForce}
\end{figure}

The combination of pressure and shear defines the normal and the tangential forces experienced by a ring of radius $\bsb$ :
\be
\label{Fn-Ft}
F_n(\bsb)\es2\pi \bsb\left( p(\bsb) + \frac{1}{2} s(\bsb)\right),
\\
F_t(\bsb)\es 2\pi \bsb \left( p(\bsb) - \frac{1}{2} s(\bsb)\right).
\ee
In Fig.~\ref{figForce} we show the plot for the normal force  $F_n$, and the tangential forces $F_t$ in the impact parameter space.
We see that the normal force  is positive. The positive nature of the normal component of the force ensures stability against collapse \cite{Polyakov:2018zvc}. 
We observe that the tangential force is positive for smaller $\bsb$ and has a negative peak at around $\bsb=0.03$~fm   ($\bsb=0.1$~fm) for $\Delta=0.2$ ($\Delta=0.5$), while for larger $\bsb$ it is zero. The presence of both positive and negative regions maintains stability in the tangential direction.

 The pressure and shear distributions for a dressed electron in QED in  Breit frame was calculated in \cite{Lorce:2018egm}. The pressure and shear distributions coming from the fermionic part of the QED EMT is negative and approaches zero near the periphery. The 2D pressure and shear distribution for the proton in the Drell-Yan frame has been calculated in \cite{Chakrabarti:2020kdc} in a light-front diquark model with inputs from AdS/QCD. The behavior is similar as seen here in the DQM. The qualitative behavior is also similar as calculated in  the light cone sum rule approach \cite{Anikin:2019kwi}  and that extracted from JLab data for the nucleon \cite{Burkert:2018bqq}. Qualitative behavior of the normal and tangential force is  similar to that observed in phenomenological models for the nucleon \cite{Lorce:2018egm,Chakrabarti:2020kdc}.      

\subsection{The energy density and pressure distributions in DQM}
We also calculate the Galilean energy density, radial pressure, tangential pressure, isotropic pressure, and pressure anisotropy, which are defined in Ref ~\cite{Lorce:2018egm} 
\be \label{Genergy}
\mu_i(\bsb) \es M \left [\frac{1}{2} A_i(\bsb)+ \overline{C}_i(\bsb) + \frac{1}{4M^2}\frac{1}{\bsb}\frac{d}{d\bsb} \left( \bsb\frac{d}{d\bsb}\left[ \frac{1}{2}B_i(\bsb) - 4C_i(\bsb) \right]\right) \right], \\
\label{radialP}
\sigma^r_i(\bsb) \es M  \left[
-\overline{C}_i(\bsb) + \frac{1}{M^2} \frac{1}{\bsb} \frac{d C_i(\bsb)}{d\bsb} 
\right],  \\
\label{tangentialP}
\sigma^t_i(\bsb) 
\es M  \left[
-\overline{C}_i(\bsb) + \frac{1}{M^2} \frac{d^2C_i(\bsb)}{d \bsbb} 
\right] \label{mp}, \\ 
\label{totalP}
\sigma_{i}(\bsb) \es M  \left[
-\overline{C}_i(\bsb) + \frac{1}{2}\frac{1}{M^2} \frac{1}{\bsb} \frac{d}{d\bsb}\left(\bsb \frac{d \ C_i(\bsb)}{d \bsb} \right)
\right] , \\
\label{shearlike}
\Pi_i(\bsb)\es  M  \left[ 
-\frac{1}{M^2} \bsb \frac{d}{d\bsb}\left(\frac{1}{\bsb} \frac{dC_i(\bsb)}{d \bsb} \right)\right].
\ee

In this section, we study five 2-dimensional distributions in impact parameter space $\bsb$. We first take the Fourier transform of all the GFFs from momentum space to impact parameter space as defined in Eq. \ref{fgff}. For all the two-dimensional distributions we choose a suitable value for the Gaussian width $\Delta=0.2$. 

The radial pressure, tangential pressure and isotropic pressure are related to the form factor $C$ and $\overline{C}$ only. The relation between radial pressure, tangential pressure and isotropic pressure is evident from Eqs. \ref{radialP} to \ref{totalP} 
\be\label{pressurerelation}
\sigma_\Q\es \frac{(\sigma^r_\Q+\sigma^t_\Q)}{2},
\ee
We can also infer from Eq. \ref{radialP}, \ref{tangentialP} and \ref{shearlike} the relation between pressure anisotropy, radial pressure and tangential pressure as
\be
\label{shearrelation}
\Pi_\Q\es \sigma^r_\Q-\sigma^t_\Q.
\ee
Therefore, from Eq. \ref{pressurerelation} we can corroborate that three pressure must show similar behavior. 
One can clearly anticipate that from Eq. \ref{shearrelation} that the nature of pressure anisotropy will have a bell shape. 

\subsection{Numerical Analysis: The energy density and pressure distributions} 
In this section we discuss, the distributions listed in Eqs. \ref{Genergy}-\ref{shearlike}. 
\begin{figure}[h]
		\includegraphics[scale=0.3]{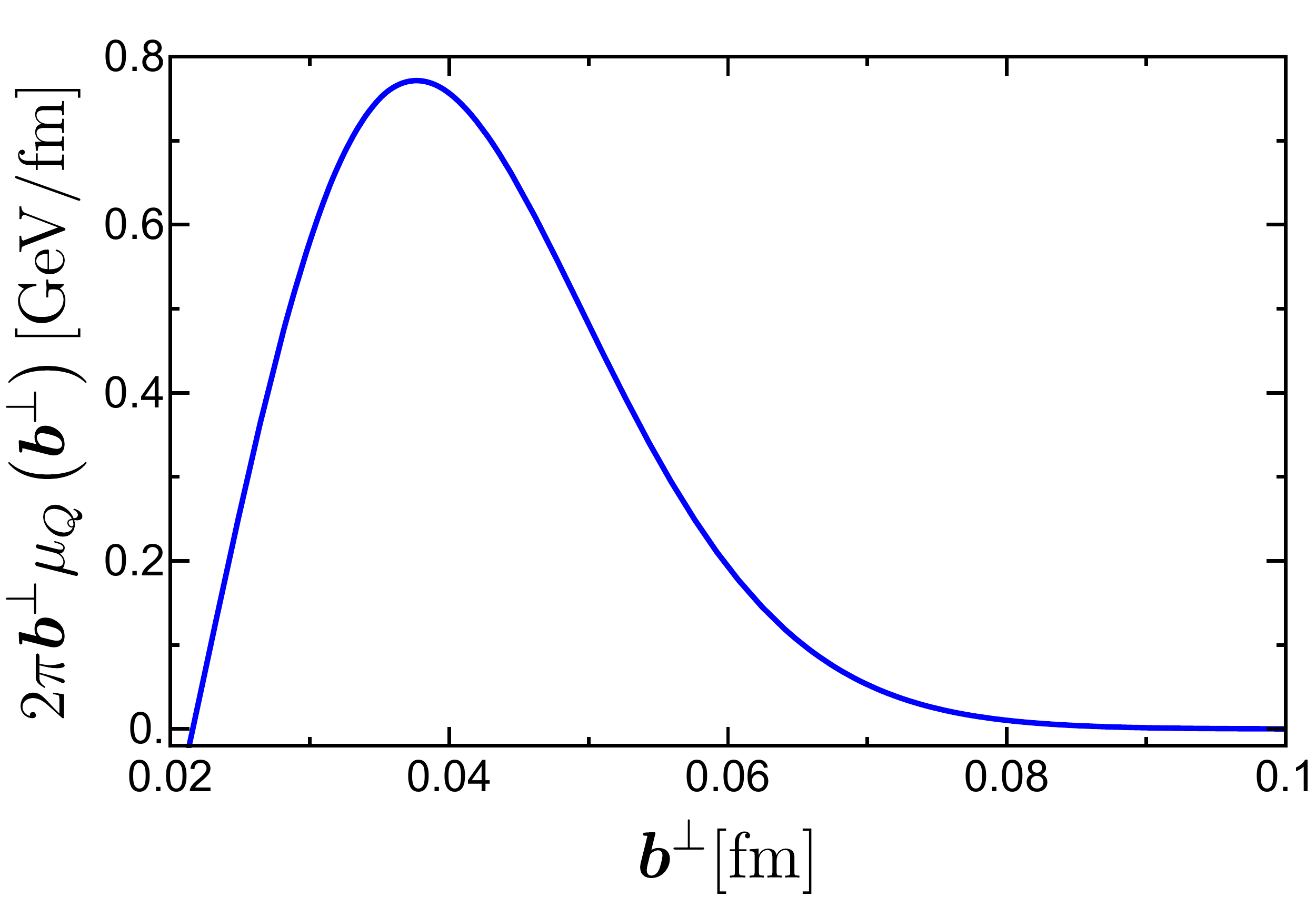}
	\caption{Plot the 2D Galilean energy density $(2\pi \bsb \mu_\Q (\bsb))$ as a function of $\bsb$.
We choose the Gaussian width $\Delta=0.2$.}
\label{energydensity}
\end{figure}
The Galilean energy density is nothing but the combination of diagonal components (from the way we extract the GFFs) of the EMT which encapsulates all the GFFs that can be seen from Eqs. \ref{rhsA} - \ref{rhsCbar} and Eq. \ref{Genergy}. In Fig \ref{energydensity}, we observe that the Galilean energy density increases with a peak value around $\bsb = 0.04~ \mathrm{fm}$ and then decreases at the periphery. 

\begin{figure}[h]
	\includegraphics[scale=0.3]{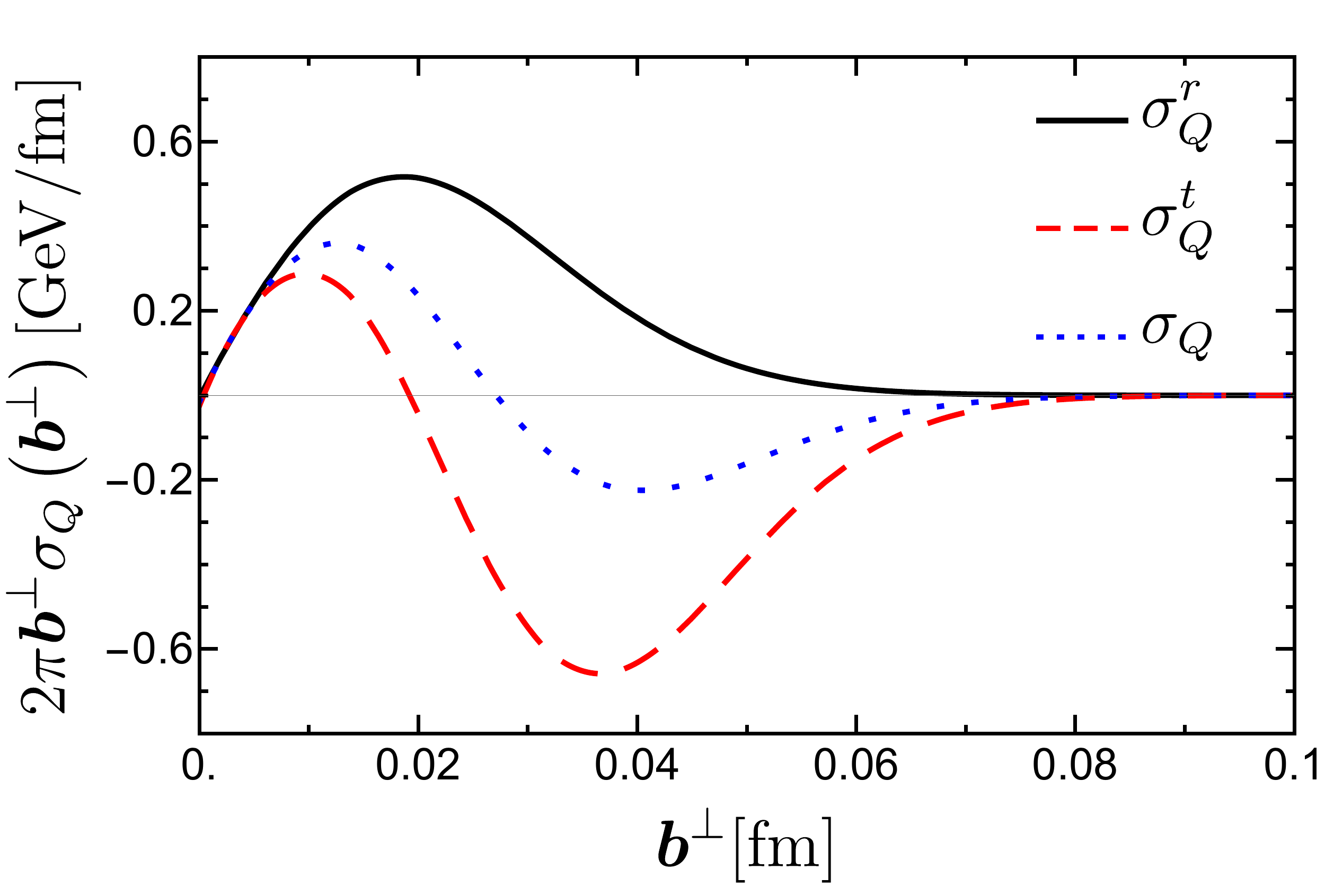}
	\caption{Plot of the 2D radial pressure $ (2\pi \bsb \sigma^r_\Q$), tangential pressure  $(2\pi \bsb \sigma^t_\Q)$,  isotropic pressure  $(2\pi \bsb \sigma_\Q)$, as a function of $\bsb$. We choose the Gaussian width $\Delta=0.2$.}
	\label{allpressure}
\end{figure}
\begin{figure}[h]
		\includegraphics[scale=0.3]{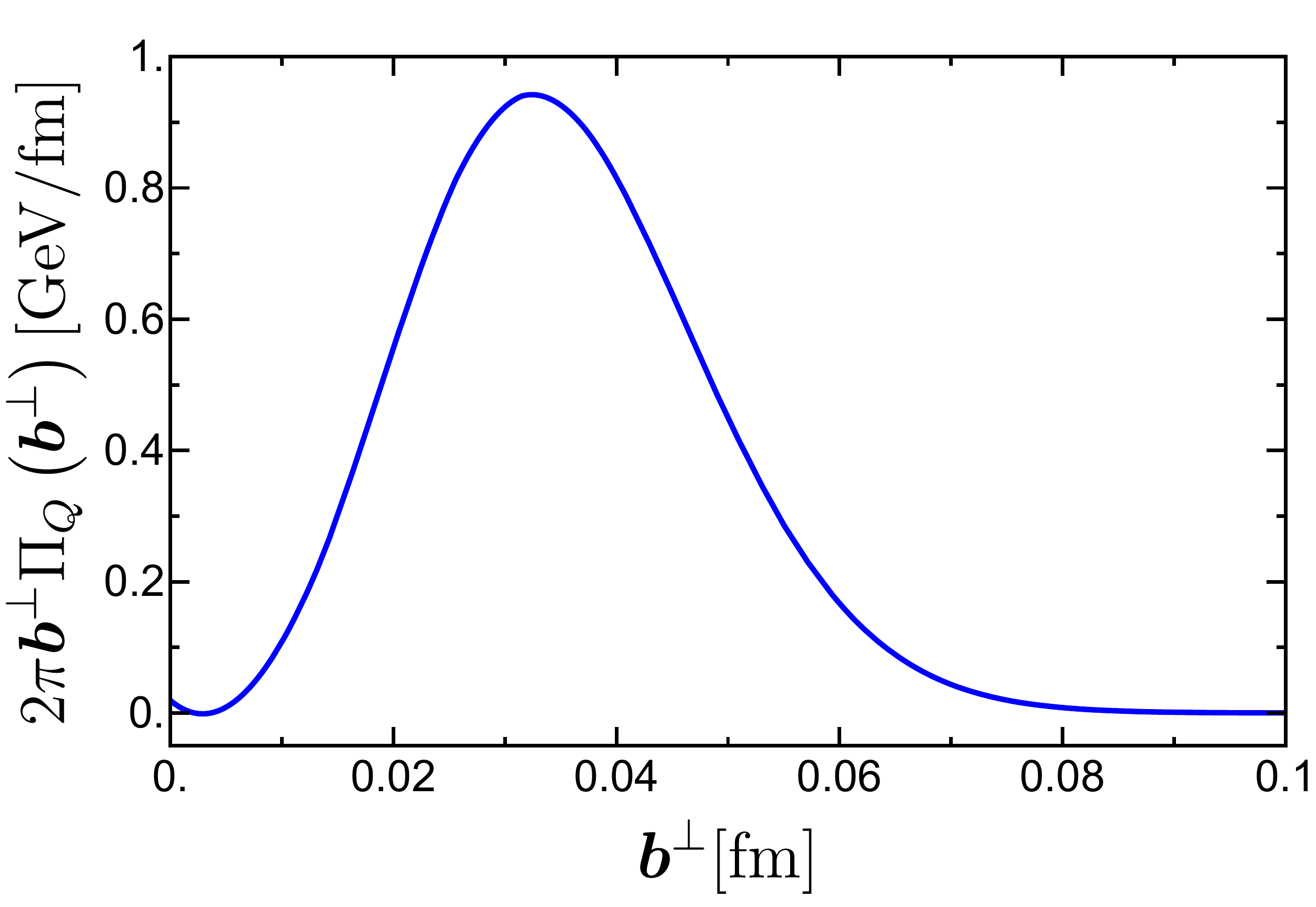}
	\caption{Plot of the 2D pressure anisotropy $(2\pi \bsb \Pi_{\Q})$ as a function of $\bsb$. We choose the Gaussian width $\Delta=0.2$.}\label{anisotropy}
\end{figure}
The energy density shows unphysical behavior (negative) for  $\bsb \le 0.02\ \mathrm{fm}$, which is different from phenomenological models for the nucleon \cite{Lorce:2018egm, Chakrabarti:2020kdc}. 
The tangential pressure flips the sign slightly away from the center. The total tangential force near the central region up to about $0.02$ fm is repulsive but becomes attractive after that till the outer region. This gives the signature that {\it quark contribution} dominates in the central region while the {\it gluon contribution} can be manifested in the small negative region away from the central region towards the periphery \cite{Lorce:2018egm}. 

The radial pressure is always positive and is governed by the quark contribution. The isotropic pressure shows a slight negative region. 
In Fig. \ref{anisotropy}, we show the plot of pressure anisotropy. Typically, it is expected that the total pressure anisotropy should vanish at the center that is one of the reasons for the radial pressure to be greater than tangential pressure. However, in our model, for the quark case, the pressure anisotropy shows a slight negative contribution around the central region. To understand this, it will be interesting to study the effect of the gluonic part of the EMT. Then, it becomes zero slightly away from the center and further, we observe that anisotropy increases monotonically till about $0.03$ fm and then again decreases. Overall, for a broad range of $\bsb$,  the 2D pressure anisotropy behaves in a similar way as in a diquark model and in a multipole model for the nucleon \cite{Lorce:2018egm}.

\section{Conclusion}
\label{con}

In this work we have investigated the GFFs and the mechanical properties like the pressure, shear and the energy distributions for a composite spin $1/2$ system, namely a quark dressed with a gluon at one loop in QCD. The GFFs and the mechanical properties of the proton in terms of its quark and gluon content have attracted a lot of interest  in recent days. There are a lot of theoretical studies in phenomenological models of the nucleon, there are also lattice calculations. Most of these theoretical models for the nucleon do not include gluons. The C form factor, or the D-term, that gives the pressure distribution inside the nucleon, in particular, is related to the non-light-cone (plus)  component of the energy-momentum tensor, and  therefore depends on the quark-gluon interaction.  In this work we replace the nucleon state by a composite spin-$1/2$ state, namely a quark dressed with a gluon at one loop in QCD, in order to calculate the GFFs. This may be thought of as a perturbative model that incorporates quark-gluon interaction. Our approach based on light-front Hamiltonian perturbation theory treats such a state fully relativistically. The LFWFs of the state   can be calculated analytically in perturbation theory, and the GFFs are expressed in terms of overlaps of LFWFs. We have used the LF two-component formalism to eliminate the constrained fields in light cone gauge.     In order to verify the correctness of the approach, we have compared with the existing results for the D-term for a dressed electron at one loop, in QED.  The pressure and shear distributions as well as the 2D energy density are calculated in the Drell-Yan frame when there is no momentum transfer in the longitudinal direction. We have compared our results with the results for the nucleon calculated in phenomenological models. Overall we have observed a qualitatively similar behavior, however the magnitude of the results in DQM  cannot be compared with that of a nucleon, which is expected. In this work we presented the GFF and the mechanical properties due to the fermionic part of the EMT. An interesting study would be to calculate the contributions coming from the gluonic part, this is part of ongoing work.     
	
\appendix
\section{GFFs in terms of LFWFs}\label{appA}	
The matrix elements in terms of overlap of LFWFs used to calculate the GFFs up-to $\mathcal{O}(g^2)$ are as follows:

\be
\mathcal{M}^{\mu\nu}_{SS'} \es \sum_{\lambda_2 \lambda'_1} \! g \int \frac{[x \bska]}{\sqrt{2(2 \pi)^3(1-x)}} \left[
\psi_1^* 
\chi_{S'}^{\dagger}\mathcal{O}_1^{\mu \nu}\chi_{\lambda'_1}
\phi^{S}_{\lambda'_1,\lambda_2}(x,\bska)\right]\nnn
\ps \sum_{\lambda_2 \lambda'_1} \! g \int\frac{[x \bska]}{\sqrt{2(2 \pi)^3(1-x)}} \left[\phi^{*S}_{\lambda'_1,\lambda_2}(x,\bska)
\chi_{\lambda'}^{\dagger}\mathcal{O}_2^{\mu \nu}\chi_{S}
\psi_1\right]\nnn
\ps  \sum_{\lambda_1 \lambda_2 \lambda'_1} \! \int [x \bska] \left[
\phi^{*S'}_{\lambda_1,\lambda_2}(x,\bskapr)
\chi_{\lambda'_1}^{\dagger}\mathcal{O}_3^{\mu \nu}\chi_{\lambda'_1}
\phi^{S'}_{\lambda'_1,\lambda_2}(x,\bska)\right],
\ee
where 
\be
 [x \bska] \es \frac{dx ~d^2\bska }{8 \pi^3}, \\
 \mathcal{O}_1^{++}\es \mathcal{O}_2^{++}=0,\\
\mathcal{O}_3^{++}\es 2 P^{+2}\  x,\\
\mathcal{O}_1^{ij}\es \frac{\left(\tilde{\sigma}^iq^j\right)\left(\tilde{\sigma}^{\perp}\cdot \epsilon_{\lambda_2}^{\perp}\right)}{4}+\frac{\left(\tilde{\sigma}^{\perp}\cdot \epsilon_{\lambda_2}^{\perp}\right)\left(\tilde{\sigma}^i(2\kappa^j+q^j)\right)}{4x}+\frac{\left(\tilde{\sigma}^i\epsilon_{\lambda_2}^j\right)\left(\tilde{\sigma}^{\perp}\cdot \bska+im\right)}{2x}+\frac{\left(\tilde{\sigma}^{\perp}\cdot \bsq-im\right)\left(\tilde{\sigma}^i\epsilon_{\lambda_2}^j\right)}{2},  \\
\mathcal{O}_2^{ij}\es  \frac{[\tilde{\sigma}^i(2\kappa^{j}-(1-2x)q^j)]\left(\tilde{\sigma}^{\perp}\cdot \epsilon_{\lambda_2}^{\perp*}\right)}{4x}+\frac{\left(\tilde{\sigma}^{\perp}\cdot \epsilon_{\lambda_2}^{\perp*}\right)\left(\tilde{\sigma}^iq^j\right)}{4}+\frac{\left(\tilde{\sigma}^{\perp}\cdot (\bska+x \bsq)-i (1-x)\ m\right)\left(\tilde{\sigma}^i\epsilon_{\lambda_2}^{j*}\right)}{2x},  \\
\mathcal{O}_3^{ij}\es \tilde{\sigma}^i ( 2\kappa^j+ q^{j})\left(\tilde{\sigma}^{\perp}\cdot\bska \right) + \tilde{\sigma}^{\perp}\cdot(2 \bska +  \bsq)\ 
\tilde{\sigma}^i (2 k^j + q^{j}).
 \ee 
In the following subsections, the diagonal (non-diagonal) contribution is indicated by subscript $\mathrm D$ ($\mathrm ND$).
\subsection{Extraction of $A_\Q(q^2)$:}
The diagonal  contribution from the one-particle sector is given by
	\begin{align}	&\left[\mathcal{M}^{++}_{\uparrow \uparrow} + \mathcal{M}^{++}_{\downarrow \downarrow}\right]_{\mathrm{1,\,D}}=2P^{+2}|\psi_{1}|^2.
	\end{align}	
The diagonal  contribution from the two-particle sector is given by
\be
\left[\mathcal{M}^{++}_{\uparrow \uparrow} + \mathcal{M}^{++}_{\downarrow \downarrow}\right]_{\mathrm{2,\,D}} \es 2 P^{+2}\  g^2\ C_F  \int [x \bska] \frac{x}{(1-x)} \frac{\left[m^2(1-x)^4+(1+x^2)(\bskasq +(1-x)\bska \cdot \bsq)\right]}{D_1  \ D_2}|\psi_{1}|^2.
\ee
The total non-diagonal contribution comes from the overlap of one particle and two-particle sectors. Here it vanishes : 
\be
\left[\mathcal{M}^{++}_{\uparrow \uparrow} + \mathcal{M}^{++}_{\downarrow \downarrow}\right]_{\mathrm{ND}} \es 0.
\ee
\subsection{Extraction of $B_\Q(q^2)$:}
The single particle sector does not contribute:
\begin{align}
	\left[\mathcal{M}^{++}_{\uparrow \downarrow} + \mathcal{M}^{++}_{\downarrow \uparrow}\right]_{\mathrm{1,\,D}}=0.
	\end{align}
From the two-particle sector there are diagonal contributions:
\be
\left[\mathcal{M}^{++}_{\uparrow \downarrow} + \mathcal{M}^{++}_{\downarrow \uparrow}\right]_{\mathrm{2,\,D}} \es  iq^{(2)}P^{+2}2 g^2\ C_F \int [x \bska]  \frac{mx^2(1-x)^2}{D_1 \ D_2}.
\ee
The total non-diagonal contribution from the overlap of one particle and two-particle sectors vanishes:
\be
\left[\mathcal{M}^{++}_{\uparrow \downarrow} + \mathcal{M}^{++}_{\downarrow \uparrow}\right]_{\mathrm{ND}} \es 0.
\ee
\subsection{Extraction of $C_\Q(q^2)$ and $\overline{C}_\Q(q^2)$:}

The single particle sector does not contribute:
\begin{align}
	\left[\mathcal{M}^{11}_{\uparrow \downarrow} + \mathcal{M}^{11}_{\downarrow \uparrow}\right]_{\mathrm{1,\,D}}=0.
	\end{align}
Contributions from the two-particle sector can be diagonal or non-diagonal:	
\be
\left[\mathcal{M}^{11}_{\uparrow \downarrow} + \mathcal{M}^{11}_{\downarrow \uparrow}\right]_{\mathrm{2,\,D}} 
\es ig^2 C_F \ q^{(2)} \int [x \bska]\frac{m(1-x)(2\kappa^{(1)}+q^{(1)})\left[(2-x)\kappa^{(1)}+(1-x)q^{(1)}\right]}{D_1\ D_2},\\
\left[\mathcal{M}^{11}_{\uparrow \downarrow} + \mathcal{M}^{11}_{\downarrow \uparrow}\right]_{\mathrm{ND}} 
\es -ig^2 C_F \ q^{(2)} \int [x \bska]  \frac{ m(1-x)}{D_1}.
\ee
Similarly for the other component 
\be
	\left[\mathcal{M}^{22}_{\uparrow \downarrow} + \mathcal{M}^{22}_{\downarrow \uparrow}\right]_{\mathrm{1,\,D}}\es0,\\
\left[\mathcal{M}^{22}_{\uparrow \downarrow} + \mathcal{M}^{22}_{\downarrow \uparrow}\right]_{\mathrm{2,D}} \es -\frac{ig^2C_F}{2}\! \!\int\!\! [x \bska]\frac{m(x-1)(2\kappa^{(2)}+q^{(2)})\left[(2\kappa^{(1)}+(1-x)q^{(1)})\ q^{(1)}-(2\kappa^{(2)}+q^{(2)})\ q^{(2)}\right]}{D_1\ D_2},
\\
\left[\mathcal{M}^{22}_{\uparrow \downarrow} + \mathcal{M}^{22}_{\downarrow \uparrow}\right]_{\mathrm{ND}} \es 0,
\ee
where 
\be
    |\psi_{1}|^2 \es 1- g^2 \ C_F\int [x \bska] \frac{\frac{\bskasq(1+x^2)}{(1-x)}+m^2(1-x)^3}{D_1^2},\\D_1:\es \bskasq +m^2(1-x)^2, \\D_2:\es\left(\bska +(1-x)\bsq\right)^2+m^2(1-x)^2.
\ee

\section{Integrals used to calculate GFFs}\label{appB}	

The following integrals are used to calculate the analytical forms of the GFFs

\be
	\int d^2\bska \frac{1}{D_1} \es \pi \log\left[\frac{\Lambda^2+ m^2(1-x)^2}{m^2(1-x)^2}\right] ,\\
	\int d^2\bska \frac{1}{D_1 D_2} \es \frac{\pi}{(1-x)^2} \frac1{q^2}\ \frac{f_2}{ f_1} ,\\
	\int d^2\bska \frac{\kappa^{(i)}}{D_1 D_2} \es -\frac{\pi}{(1-x)}\frac{ q^{(i)}}{q^2 }\frac{f_2}{2 f_1}, \\
	\int d^2\bska \frac{\kappa^{(1)}\kappa^{(2)}}{D_1 D_2}\es \pi\frac{q^{(1)}q^{(2)}}{q^2}\bigg[-1+\left(1+\frac{2m^2}{q^2}\right)\frac{f_3}{2f_1}\bigg],\\
	\int d^2\bska\frac{(\kappa^{(i)})^2}{D_1D_2}\es\pi\bigg[-f_1 \ f_3+\frac{1}{2}+\frac{(q^{(i)})^2}{q^2}\left[\left(1+\frac{2m^2}{q^2}\right)\frac{f_3}{2f_1}-1\right]+\frac{1}{2}\log\bigg(\frac{\Lambda^2}{m^2(1-x)^2}\bigg)\bigg], \ee
where, $i= (1,2)$ and
\be 
	D_1:\es \bskasq +m^2(1-x)^2, \\D_2:\es\left(\bska +(1-x)\bsq\right)^2+m^2(1-x)^2,\\	f_1 :\es\frac{1}{2}\sqrt{1+\frac{4 m^2}{q^2}}, \\ f_2:\es\log\left(1+\frac{q^2\left(1+2f_1\right)}{2 m^2}\right), \\
	f_3:\es \log\left(\frac{1+2 f_1}{-1+2 f_1}\right)
	\ee
\section*{Acknowledgments}
JM would like to thank the Department of Science and Technology (DST), Government of India, for financial support through Grant No. SR/WOS-A/PM-6/2019(G). SN thank the Chinese Academy of Sciences Presidents International Fellowship Initiative for the
support via Grants No. 2021PM0021. AM would like to thank SERB-POWER Fellowship (SPF/2021/000102) for financial support. 
	
\bibliographystyle{elsarticle-num}
\bibliography{references}		
\end{document}